\documentclass[12pt]{iopart}
\usepackage{iopams}
\newtheorem{lemma}{Lemma}
\newtheorem{proposition}{Proposition}
\newtheorem{theorem}{Theorem}
\newtheorem{corollary}{Corollary}
\newtheorem{definition}{Definition}
\def\tr{\mathop{\rm tr}\nolimits}

\def\ci{\mathop{\textrm{i}}\nolimits}
\def\cic{\mathop{\rm i}\nolimits}

\begin{document}

\title[Vacuum type I and Papapetrou fields]
{Vacuum type I spacetimes and aligned Papapetrou fields: symmetries}

\author{Joan Josep Ferrando$^1$\
and Juan Antonio S\'aez$^2$}

\address{$^1$\ Departament d'Astronomia i Astrof\'{\i}sica, Universitat
de Val\`encia, E-46100 Burjassot, Val\`encia, Spain.}

\address{$^2$\ Departament de Matem\`atica Econ\`omico-Empressarial, Universitat de
Val\`encia, E-46071 Val\`encia, Spain}

\ead{joan.ferrando@uv.es; juan.a.saez@uv.es}

\begin{abstract}
We analyze type I vacuum solutions admitting an isometry whose
Killing 2--form is aligned with a principal bivector of the Weyl
tensor, and we show that these solutions belong to a family of
type I metrics which admit a group $G_3$ of isometries. We give a
classification of this family and we study the Bianchi type for
each class. The classes compatible with an aligned Killing 2--form
are also determined. The Szekeres-Brans theorem is extended to non
vacuum spacetimes with vanishing Cotton tensor.

\end{abstract}

\pacs{0420C, 0420-q}

\submitto{\CQG}


\section{Introduction}

Several methods have been developed to simplify Einstein equations
in order to look for new exact solutions. These approaches usually
imply imposing conditions that restrict the space of possible
solutions. Thus, a notable number of known solutions have been
obtained under the hypothesis that they admit a fixed isometry or
conformal group. It has also been fruitful to impose restrictions
on the algebraic structure of the Weyl tensor. Indeed, wide
families of algebraically special solutions of Einstein equations
have been found by considering coordinates or frames adapted to
the multiple Debever direction that these spacetimes admit.
Nevertheless, there is a lack of knowledge about algebraically
general solutions, and they have usually been obtained by imposing
spacetime symmetries. One way to correct this situation is to opt
for imposing a complementary condition on a type I Weyl tensor
which allows us to simplify Einstein equations. This means
considering subclassifications of algebraically general spacetimes
and looking for solutions in every defined class.

Debever \cite{deb} was the first to suggest a classification of
type I metrics based on the nullity of one of the Weyl algebraic
invariant scalars. A similar kind of conditions are satisfied by
the Weyl eigenvalues in the purely electric or purely magnetic
solutions \cite{mcar}, as well as in some classes defined by
McIntosh and Arianhood \cite{mcar2} where the positive or negative
real character of a Weyl invariant scalar is imposed. These
classes may also be reinterpreted in terms of some kind of
degeneracy of the four Debever directions that a type I metric
admits \cite{mcar2} \cite{fsEM}. In this latest work we have shown
that these classes contain some families of metrics which can be
interpreted as generalized purely electric or purely magnetic
spacetimes.

Besides those algebraic classifications quoted above one can
define other families of type I spacetimes by imposing conditions
on the first order Weyl concomitants. For example, in the normal
shear-free perfect fluid solutions \cite{bar} one restricts the
kinematic coefficients associated with the unique time-like Weyl
principal direction. Another classification that imposes first
order differential conditions on the type I canonical frame was
given by Edgar \cite{edgar2}. We have changed here this approach
slightly by imposing invariant conditions that are symmetric with
respect to the three space-like principal directions. When the
spacetime Cotton tensor vanishes, the Bianchi identities imply
that in both Edgar's and our degenerated classes, the Weyl
eigenvalues are functionally dependent. The present work is
concerned with the vacuum solutions of the more degenerated class
which we name $I_1$. We show that in this case a 3--dimensional
group of isometries exists and all the Weyl invariant scalars
depend on a unique real function. We also offer an invariant
subclassification of the type $I_1$ metrics and we prove that the
classes can be characterized by the Bianchi type of the $G_3$.
Some of these results were communicated without proof at the
Spanish Relativity Meeting 1998 \cite{fsERE1}.

In an algebraically general spacetime the Weyl canonical frame
determines three time-like principal 2--planes associated with a
frame of orthonormal eigenbivectors of the Weyl tensor. Here we
use the complex Cartan formalism adapted to this orthonormal frame
of eigenbivectors under the hypothesis of a vanishing Cotton
tensor. In this case the Cartan structure equations and their
integrability conditions, the Bianchi identities, do not
constitute a completely integrable system of equations. The
integrability conditions for this system were considered by Bell
and Szekeres \cite{besz} and were investigated by Brans
\cite{brans2} who named them the {\it post-Bianchi equations}.
Edgar \cite{edgar1} analyzed in detail these post-Bianchi
equations that we obtain easily here using a plain covariant
approach. On the other hand, the complex formalism that we use
allows us to generalize easily the Szekeres-Brans theorem on the
non existence of vacuum type I solutions with a zero Weyl
eigenvalue \cite{sze} \cite{brans1}. This extension is similar to
that known for the Goldberg-Sachs theorem and it implies that the
Szekeres-Brans result also applies to the non vacuum solutions
with vanishing Cotton tensor. This fact has already been used in
analyzing the restrictions on the existence of purely magnetic
solutions \cite{fsM}. It is worth pointing out that the set of the
metrics with vanishing Cotton tensor contains all the vacuum
solutions as well as other non vacuum metrics with specific
restrictions on the energy content. For example, in the case of
perfect fluid solutions, this condition leads to non-shearing
irrotational flows \cite{tau}.

When a spacetime admits an isometry, the Killing vector $\xi$
plays the role of an electromagnetic potential satisfying the
Lorentz gauge, and its covariant derivative $\nabla \xi$ is, in
the vacuum case, a solution of the source-free Maxwell equations.
Because this fact was first observed by Papapetrou \cite{pap}, the
{\it Killing 2--form} $\nabla \xi$ has also been called the {\it
Papapetrou field} \cite{faso1}. Metrics admitting an isometry were
studied by considering the algebraic properties of the associated
Killing 2--form \cite{deb1} \cite{deb2}, and this approach was
extended to the spacetimes with an homothetic motion \cite{mc1}
\cite{mc2}.

More recently Fayos and Sopuerta \cite{faso1} \cite{faso2} have
developed a formalism that improves the use of the Killing 2--form
and its underlined algebraic structure in the analysis of the
vacuum solutions with an isometry. They consider two new
viewpoints that permit a more accurate classification of these
spacetimes: (i) the differential properties of the principal
directions of the Killing 2--form, and (ii) the degree of
alignment of the principal directions of the Killing 2--form with
those of the Weyl tensor. The Fayos and Sopuerta approach uses the
Newman-Penrose formalism and several extensions have been built
for homothetic and conformal motions \cite{ste} \cite{lud} and for
non vacuum solutions \cite{faso3}.

In the Kerr geometry the principal directions of the Killing
2--form associated with the time-like Killing vector are the two
double principal null (Debever) directions of the Weyl tensor
\cite{faso1}. But in a type D spacetime these two Debever
directions are the principal directions of the principal 2--plane
and, consequently, in the Kerr spacetime, the Killing 2--form is
an eigenbivector of the Weyl tensor. This fact has also been
remarked by Mars \cite{mars} who has shown that this property
characterizes the Kerr solution under an asymptotic flatness
behavior. Elsewhere \cite{fsDB} we have shown that the type D
vacuum solutions having a time-like Killing 2--form aligned with
the Weyl geometry are the Kerr-NUT spacetimes.

In a Petrov type I spacetime a principal direction of a Weyl
principal 2--plane never coincides with one of the four null
Debever directions \cite{bel} \cite{fsI}. Therefore, in a type I
metric two different kinds of alignment of a Killing 2--form and
the Weyl tensor can be considered. On one hand, we can impose, as
Fayos and Sopuerta do \cite{faso2}, the alignment between a
principal direction of the Killing 2--form and a principal Debever
direction of the Weyl tensor. On the other hand, we can consider
that a principal 2--plane of the Weyl tensor is the principal
2--plane of the Killing 2--form. In this work we adopt this second
point of view and we show that all the vacuum type I solutions
having a Killing 2--form which is an eigenbivector of the Weyl
tensor belong to the class $I_1$ quoted above. In consequence,
these spacetimes admit a group $G_3$ of isometries. We also study
the Bianchi types compatible with this kind of alignment.

The article is organized as follows. In section 2 we present the
general notation of the complex formalism used in this work and we
apply this formalism to write the Cartan structure equations and
to characterize the geometric properties of a 2+2 almost-product
structure. In section 3 we adapt the Cartan complex formalism to
the canonical frame of a type I Weyl tensor and we write the
Bianchi identities for the spacetimes with a vanishing Cotton
tensor. Some direct consequences of these equations lead us to
present a classification of the type I spacetimes. The
generalization of a theorem by Brans and a tensorial expression
for the post-Bianchi equations are easily obtained too. In section
4 we study general properties of an aligned Papapetrou field and
we show that every type I vacuum solution with an isometry having
an aligned Killing 2--form belongs to the more degenerated class
$I_1$ in the classification given in the previous section. Some
basic results on the type $I_1$ vacuum solutions are presented in
section 5 where a subclassification of these metrics is also
offered. In section 6 we show that a type $I_1$ vacuum metric with
a non constant eigenvalue admits a group $G_3$ of isometries and,
for every subclass presented in previous section, we study its
Bianchi type. The Bianchi types which are compatible with an
aligned Killing 2--form are also determined. Finally, in section 7
we study the vacuum case with constant eigenvalues and we prove
that the spacetime has a group $G_4$ of isometries, obtaining in
this way an intrinsic characterization of the vacuum homogeneous
Petrov metric.

\section{Cartan equations in complex formalism}
Let $(V_4, g)$ be an oriented spacetime of signature $(-,+,+,+)$
and let $\eta$ be the volume element. If we denote $*$ the Hodge
dual operator, we can associate with every 2-form $F$ the
self-dual 2-form ${\cal F} = \frac{1}{\sqrt{2}} ( F - \ci *F )$.
Hereafter we shall say bivector to indicate a self-dual 2-form.

We can associate with every oriented orthonormal frame $\{
e_{\alpha} \}$ the orthonormal frame $\{ {\cal U}_i \}$ of the
bivectors space defined as $\, {\cal U}_i  = \frac{1}{\sqrt{2}}
\left[ e_0 \wedge e_i - \ci *(e_0 \wedge e_i) \right] $. This
frame satisfies $\, 2{\cal U}_i^2 = g$, and it has the induced
orientation given by
\begin{equation} \label{orient}
{\cal U}_i \times {\cal U}_j = - \frac{\ci}{ \sqrt{2}} \epsilon_{ijk} \
{\cal U}_k  \, , \qquad \quad i \neq j
\end{equation}
where $\times$ means the contraction of adjacent indexes in the tensorial
product and $\, {\cal U}_i^2 = {\cal U}_i \times {\cal U}_i$.

On the other hand, if $\{ {\cal U}_i \}$ is an oriented orthonormal
frame of bivectors, for every (complex) non null vector $x$, the four
directions $\{ x, {\cal U}_i (x) \}$ are orthogonal and can be normalized
to get an oriented orthonormal frame. In terms of these vectors, the metric
tensor $g$ writes
\begin{equation}   \label{metrica1}
g = \frac{1}{x^2} \Big[ x \otimes x - 2 \ \sum_{i=1}^3 {\cal U}_i (x)
\otimes  {\cal U}_i (x) \Big]
\end{equation}
Moreover, the self dual 2-forms ${\cal U}_i$ can also be written as
 \begin{equation}\label{ui}
{\cal U}_i = - \frac{1}{{ x}^2} \left(
{ x} \wedge {\cal U}_i (x) \ + \frac{\ci}{ \sqrt{2}} \
\epsilon_{ijk} \, {\cal U}_j (x) \wedge  {\cal U}_k  (x)
\right)
\end{equation}
Nevertheless, if $x$ is a null vector the four orthogonal directions
$\{ x ,  {\cal U}_i (x) \}$ can not be independent. So, one of the
directions is a linear combination of the other ones,
${\cal U}_i (x) = a x + b^j {\cal U}_j (x)$. Contracting this equation
with ${\cal U}_i$  we obtain the following
\begin{lemma}  \label{is}
If $\{ {\cal U}_i \}$ is an orthonormal frame of bivectors and $x$ is a
null vector, then scalars $b^i$ exist such that
$$x = \sum b^i \ {\cal U}_i (x) $$
\end{lemma}

Every unitary bivector ${\cal U}_i$ defines a 2+2 almost product
structure. These kind of structures have been considered elsewhere
\cite{fsD} and have been used to classify the type D spacetimes
attending to the geometric properties of the Weyl principal
structure.  The 1-form $\delta {\cal U}_i \equiv - \tr \nabla
{\cal U}_i$ collects the information about the minimal and the
foliation character of the 2-planes that ${\cal U}_i$ defines.
More precisely, we have \cite{fsD} \cite{fsKY}:
\begin{lemma} \label{minifoli}
Let ${\cal U}_i $ be a unitary bivector and $\lambda_i = {\cal U}_i (\delta {\cal U}_i )$,
and let us consider the 2+2 almost product structure that ${\cal U}_i$ defines. It holds:

i) Both planes are minimal if, and only if, $Re(\lambda_i) =0$.

ii) Both planes are foliations if, and only if, $Im(\lambda_i)=0$.
\end{lemma}
The umbilical nature of the 2-planes defined by ${\cal U}_i$ can also be characterized in
terms of the covariant derivative of ${\cal U}_i$ \cite{fsD} \cite{fsKY}. This property is
equivalent to the null principal directions of ${\cal U}_i$ to
be shear-free geodesics and can be stated in terms of the 1-forms $\lambda_j$ as:
\begin{lemma} \label{umbilical}
Let us consider the 2+2 structure defined by a bivector ${\cal U}_1$, and let us take $\{
{\cal U}_2 , {\cal U}_3 \}$ to complete an orthonormal frame. The principal directions of
 $\ {\cal U}_1$ are shear-free geodesics if, and only if, $\lambda_2 = \lambda_3$.
\end{lemma}

For a given orthonormal frame $\{ e_{\alpha} \}$ six connection 1-forms
$\omega_{\alpha}^{\beta}$ ($\omega_{\alpha}^{\beta}=-\omega_{\beta}^{\alpha}$)  are defined
by $\nabla e_{\alpha} = \omega_{\alpha}^{\beta} \otimes  e_{\beta}$. These equations are
equivalent to the first  structure equations in the Cartan formalism.
If we consider $\{ {\cal U}_i \}$ the orthonormal frame of bivectors
associated with $\{ e_{\alpha} \}$, the six connection 1-forms can be
collected into three complex ones $\Gamma_i^j$ ($\Gamma_i^j = - \Gamma_j^i$)
that are defined by
$$\Gamma_i^j = \omega_i^j - \ci \ \epsilon_{ijk} \ \omega_0^k $$
In terms of these complex 1-forms, the first  Cartan structure equations
are equivalent to
\begin{equation}\label{es1a}
\nabla {\cal U}_i = \Gamma_i^j \otimes {\cal U}_j
\end{equation}
The second Cartan structure equations follow by applying the Ricci
identities to the bivectors $\{ {\cal U}_i \}$, $\nabla_{[\alpha}
\nabla_{\beta]} {{\cal U}_{i}}_{\, \epsilon \delta} = {{{\cal
U}_{i}}_{\, \epsilon}}^{\mu} R_{\mu \delta  \beta \alpha} + {{\cal
U}_{i}^{\ \mu}}_{\delta}  R_{\mu \epsilon  \beta \alpha} \, $, and
they can be written as
\begin{equation} \label{es2a}
\mbox{d} \Gamma_i^k -\Gamma_i^j \wedge \Gamma_j^k  = \ci \sqrt{2}
\epsilon_{ikm} \mbox{Riem}({\cal U}_m)
\end{equation}
where ${Riem({\cal U}_m)}_{\alpha \beta}  = \frac{1}{2} R_{\alpha \beta \mu
\nu} ({\cal U}_m)^{\mu \nu} $.
Taking into account the invariant decomposition of the Riemann tensor into the
trace--free part, the Weyl tensor, and the trace, the Ricci tensor, we have
\begin{equation*}
Riem =W  + Q \wedge g , \qquad  \qquad
 Q=\frac{1}{2} [Ric -
\frac{1}{6}(\mbox{tr}Ric) g ]
\end{equation*}
As the self-dual Weyl tensor ${\cal W} =\frac{1}{2} ( W - \ci * W)$ satisfies
${\cal W}({\cal U}) =W({\cal U})$ for every self-dual 2-form ${\cal U}$, the
second term of equations (\ref{es2a}) writes:
\begin{equation} \label{rieman}
Riem({\cal U}_m) = {\cal W}({\cal U}_m ) + {\cal U}_m \times Q + Q \times {\cal U}_m
\end{equation}

The three complex 1-forms $\lambda_i = {\cal U}_i (\delta {\cal U}_i )$
contain the 24 independent connection coefficients as the $\Gamma_i^j$ do.
In fact, by using (\ref{orient}) and the first structure equations (\ref{es1a}),
both sets $\{ \Gamma_i^j \}$ and $\{ \lambda_i \}$ can be related by
\begin{equation} \label{05}
\lambda_i \equiv {\cal U}_i (\delta {\cal U}_i ) = -\frac{\ci }{\sqrt{2}}\
\epsilon_{ijk} \ {\cal U}_k (\Gamma_i^j )
\end{equation}
And the inverse of these expressions say that for $i, j , k$ different
\begin{equation}\label{cap3123}
 {\cal U}_k(\Gamma_i^j )  = \frac{\ci}{\sqrt{2}} \, \epsilon_{ijk} \,
(  \lambda_i +\lambda_j-\lambda_k ) \, , \qquad (i, j , k \ \neq )
\end{equation}
There is a subset of the second structure equations (\ref{es2a})
that can be concisely stated in terms of the 1-forms $\lambda_i$.
Indeed, if we calculate $(\mbox{d} \Gamma_i^j, {\cal U}_k)$ from
the second structure equations (\ref{es2a}) and use (\ref{rieman})
to replace $Riem({\cal U}_k , {\cal U}_k)$, we get  that for
$i,j,k$ different, it holds:
\begin{equation}\label{traza}
\nabla \cdot  \lambda_i = \lambda_{i}^2  -  (\lambda_{j}-
\lambda_{k})^2 -  {\cal W}({\cal U}_i , {\cal U}_i ) - \frac{1}{2} \ \mbox{tr} Q,
\qquad (i, j , k \ \neq )
\end{equation}
where we have denoted $\nabla \cdot \equiv \tr \nabla $ and
$\lambda_i^2 = g(\lambda_i , \lambda_i)$.

\section{Type I metrics with vanishing Cotton tensor: generalized Szekeres-Brans theorem}

In a type I spacetime the Weyl tensor defines an orthonormal frame $\{ {\cal U}_i \}$
of eigenbivectors. If we denote $\alpha_i$ the corresponding eigenvalues, the self-dual
Weyl tensor takes the canonical expression \cite{fms}:
\begin{equation} \label{can1}
{\cal W} = - \sum \alpha_i  \ {\cal U}_i  \otimes {\cal U}_i
\end{equation}
So, the structure equations (\ref{es1a}), (\ref{es2a}) can be
written in this frame of principal bivectors. The integrability
conditions for the second structure equations (\ref{es2a}) are the
Bianchi identities which equal the divergence of the Weyl tensor
with the Cotton tensor $P$:
\begin{equation} \label{bianchi0}
\nabla \cdot W = P \, , \qquad \quad P_{\mu \nu , \beta} \equiv
\nabla_{[\mu} Q_{\nu] \beta}
\end{equation}
Hereafter we will consider spacetimes with vanishing Cotton
tensor. Then, the Bianchi identities (\ref{bianchi0}) state that
the Weyl tensor is divergence-free, $\nabla \cdot {\cal W} =0$. If
we use the canonical expression (\ref{can1}) to compute the
divergence of the Weyl tensor and we take into account that
$*{\cal U}_j = \ci {\cal U}_j$, the Bianchi identities write
\begin{equation} \label{bianchia}
\mbox{d} \alpha_i = (\alpha_j - \alpha_k )  \ (\lambda_j -
\lambda_k )- 3 \ \alpha_i \ \lambda_i  \, , \qquad (i, j , k \
\neq )
\end{equation}
where $i, j, k$ take different values. Let us suppose now that one
of the eigenvalues, say $\alpha_i$, takes the value zero. So, from
(\ref{bianchia}) we have $\, 0= (\alpha_j - \alpha_k ) \
(\lambda_j - \lambda_k)$. As $\alpha_j \neq \alpha_k$, it must be
$\lambda_j - \lambda_k =0$. This condition is equivalent to the
principal directions of ${\cal U}_i$ being shear-free geodesics as
Lemma \ref{umbilical} states, and by the generalized
Goldberg-Sachs theorem \cite{kra} the spacetime must be
algebraically special.  Thus we have generalized a previous
result, which can be inferred from a paper by Szekeres \cite{sze}
and was stated by Brans \cite{brans1}, on the non existence of
vacuum type I solutions with a vanishing Weyl eigenvalue to the
case of non vacuum solutions with vanishing Cotton tensor.
\begin{theorem}
There is no type I spacetime with vanishing Cotton tensor for
which one of the eigenvalues of the Weyl tensor is zero.
\end{theorem}

The Weyl tensor is trace-free, $\, \alpha_1 +  \alpha_2 + \alpha_3
= 0 \,$, and consequently only two of the three equations
(\ref{bianchia}) are independent. So, we can write Bianchi
identities explicitly as
\begin{equation}\label{bianchi}
\begin{array}{l}
\displaystyle \mbox{d} \alpha_1 = (\alpha_1 + 2 \alpha_2) (
\lambda_2 - \lambda_3) - 3
\alpha_1 \lambda_1 \\[3mm]
\displaystyle \mbox{d} \alpha_2 = (2\alpha_1 + \alpha_2) ( \lambda_1 - \lambda_3) -
3 \alpha_2 \lambda_2
\end{array}
\end{equation}
From these equations a direct calculation leads us to the following:
\begin{proposition} \label{anterior}
In a type I spacetime with vanishing Cotton tensor the scalars
$\{\alpha_i\}$ depend on a function ($\mbox{\rm d} \alpha_i \wedge
\mbox{\rm d} \alpha_j =0$) if, and only if, one of the following
conditions hold:

i) $\lambda_i \wedge \lambda_j =0$, $\forall i, j$,

ii) $\sum p^i \ \lambda_i =0$, where the scalars $p^i$ satisfy
$\displaystyle \sum_{i,j,k \neq}
p^i \ (\alpha_j - \alpha_k)^2 =0$.
\end{proposition}
If we take into account lemmas \ref{minifoli} and \ref{umbilical},
the last proposition states that the functional dependence of the
Weyl eigenvalues is related with restrictions on the geometric
properties of the principal 2--planes. Of course we are under the
hypothesis of a vanishing Cotton tensor. We can find a similar
situation in type D spacetimes where some families determined by
imposing conditions on the gradient of the Weyl eigenvalue turn
out to be those classes defined attending the geometric properties
of the Weyl principal structure \cite{fsD}. The above result for
the case of type I metrics leads us to the following
classification \cite{fsERE1}:
\begin{definition} \label{clas1}
We will say that a type I spacetime is of class I$_a$ ($a =1,2,3$)
if the dimension of the space that the $\lambda_i$ generate is
$a$.
\end{definition}
Differential conditions of this kind were imposed by Edgar
\cite{edgar1} on the type I spacetimes, and he showed that in the
vacuum case his classification also has consequences on the
functional dependence of the Weyl eigenvalues. We have slightly
modified the Edgar approach in order to obtain a classification
that is symmetric in the principal structures of the Weyl tensor.
We stress the invariant nature of this classification. It is based
on the vector Weyl invariants $\lambda_i$ which have a precise
geometric meaning: they contain the information about the
properties of the Weyl principal planes (see lemmas 2 and 3). On
the other hand, these geometric properties can be interpreted in
terms of the kinematical behavior of the null principal directions
of these planes \cite{fsD}. Let us take into account that
$\lambda_i$ can not be all zero because this fact implies that all
of the connection coefficients are so, and the spacetime would be
plane. So, after definition \ref{clas1}, proposition
\ref{anterior} can be stated as:
\begin{proposition}
In a type I spacetime with vanishing Cotton tensor the Weyl
eigenvalues depend on a function ($\mbox{\rm d} \alpha_i \wedge
\mbox{\rm d} \alpha_j =0$) if, and only if, it is class I$_1$ or
it is class I$_2$ and the second condition of proposition
\ref{anterior} is satisfied.
\end{proposition}

Different authors \cite{brans1} \cite{edgar1} have shown that when
the Cartan structure equations in vacuum are referred to the Weyl
principal frame the Bianchi identities have non trivial
integrability conditions. First considered by Bell and Szekeres
\cite{besz}, these integrability conditions were called the
post-Bianchi equations \cite{brans1} and they have usually been
written in the NP formalism \cite{brans1} \cite{edgar1}. Here we
can easily obtain these post-Bianchi equations in tensorial
formalism for spacetimes with vanishing Cotton tensor. Indeed,
taking the exterior derivative of (\ref{bianchi}) we obtain
\begin{equation}\label{pbianchi}
\begin{array}{l}
   \hspace{-1.5cm} \frac{1}{\alpha_2 - \alpha_3 } \mbox{d} \lambda_1 +
\frac{1}{\alpha_3 - \alpha_1} \mbox{d} \lambda_2 +
\frac{1}{\alpha_1 - \alpha_2 } \mbox{d} \lambda_3 =0 \\[3mm]
 \hspace{-1.5cm} \frac{\alpha_1}{\alpha_2 - \alpha_3 } \mbox{d} \lambda_1 +
\frac{\alpha_2}{\alpha_3 - \alpha_1} \mbox{d} \lambda_2 +
\frac{\alpha_3}{\alpha_1 - \alpha_2 } \mbox{d} \lambda_3 +
4 (\lambda_1 \wedge \lambda_2+ \lambda_2 \wedge \lambda_3 + \lambda_3 \wedge \lambda_1) =0
\end{array}
\end{equation}
If we take into account the Bianchi identities (\ref{bianchi}),
the integrability conditions of the post-Bianchi equations
(\ref{pbianchi}) are now an identity and so, no new equations are
obtained  as other authors claimed \cite{brans1} \cite{edgar1}.
Moreover, only 9 of these 12 complex equations are independent of
the Cartan structure equations \cite{brans1}.

\section{Aligned Papapetrou fields. The vacuum type I case}

If $\xi$ is a (real) Killing vector its covariant derivative
$\nabla \xi$ is named {\it Killing 2--form} or {\it Papapetrou
field} \cite{pap} \cite{faso1}. The Papapetrou fields have been
used to study and classify spacetimes admitting an isometry or an
homothetic or conformal motion (see references \cite{faso1} to
\cite{faso3}). In this way, some classes of vacuum solutions with
a principal direction of the Papapetrou field aligned with a
(Debever) null principal direction of the Weyl tensor have been
considered \cite{faso2}. Also, the alignment between the Weyl
principal plane and the Papapetrou field associated with the
time-like Killing vector has been shown in the Kerr geometry
\cite{faso2} \cite{mars}.

Is it possible to determine all the vacuum solutions having this
property of the Kerr metric? Elsewhere \cite{fsDB} we give an
affirmative answer to this question for the case of type D
spacetimes by showing that {\it the type D vacuum solutions with a
time-like Killing 2--form aligned with the Weyl geometry are the
Kerr-NUT metrics}. Here we analyze the type I case and we will see
that the spacetime necessarily admits a group $G_3$ of isometries.
We begin by showing in this section that these solutions belong to
class $I_1$ of definition \ref{clas1}.

If $\{ {\cal U}_i \}$ is an orthonormal basis of the self-dual
2-forms space, the Papapetrou field $\nabla \xi$ writes:
\begin{equation} \label{ka1}
\nabla \xi = \sum \Omega_i {\cal U}_i + \sum \tilde{\Omega}_i \tilde{\cal{U}}_i
\end{equation}
where $\Omega_i$ are three complex functions and $\tilde{}$ means
complex conjugate. Let us suppose that ${\cal U}_1$ is
$\xi$-invariant, that is, ${\cal L}_{\xi} {\cal U}_1 =0$. If we
denote $A^t$ the transposed of the tensor $A$, this condition
reads:
\begin{equation}  \label{ka2}
i(\xi) \nabla {\cal U}_1 + (\nabla \xi \times {\cal U}_1 ) - (\nabla \xi \times {\cal U}_1)^t =0
\end{equation}
Contracting this equation with ${\cal U}_2$ and ${\cal U}_3$, we get
\begin{equation} \label{ka3}
\Omega_2 = -\frac{\ci}{\sqrt{2}}(\xi , \Gamma_3^1 ) , \qquad  \quad
\Omega_3 = -\frac{\ci}{\sqrt{2}}(\xi , \Gamma_1^2 )
\end{equation}
So, if ${\cal U}_1$ is $\xi$-invariant, two complex components (or
four real ones) of the Killing 2--form $\nabla \xi$  are
determined by $\xi$. If, in addition, ${\cal U}_2$ (and so ${\cal
U}_3$) are invariant, then $\nabla \xi$ is totally determined by
$\xi$. As a consequence of this result, a group that acts on a
spacetime admitting an invariant frame must be simply transitive.
In a type I spacetime the Weyl tensor defines an invariant
orthonormal frame $\{ {\cal U}_i \}$ of eigenbivectors. Thus, we
have the following:
\begin{proposition} \label{prop-al-1}
Let $\{ {\cal U}_i \}$ be the principal 2-forms of a type I
spacetime and $\{ \Gamma_i^j \}$ the associated complex connection
1-forms. If $\xi$ is a Killing field, then the Papapetrou field
writes $\nabla \xi = \sum \Omega_i {\cal U}_i + \sum
\tilde{\Omega}_i \tilde{\cal{U}}_i $ where, for every cyclic
permutation,
\begin{equation} \label{ka4}
\Omega_i \equiv - (\nabla \xi , {\cal U}_i ) = - \frac{\cic}{\sqrt{2}}(\xi , \Gamma_j^k)
\end{equation}
\end{proposition}

In order to clarify what kind of alignment between the Killing
2--form and the Weyl tensor is analyzed in this work we give the
following:
\begin{definition}
We say that a Papapetrou field $\nabla \xi$ is aligned with a
bivector ${\cal U}$ if both 2--forms have the same principal
2--planes, that is, $\nabla \xi = \Omega {\cal U} + \tilde{\Omega}
\tilde{\cal U}$. \\
We say that a Papapetrou field $\nabla \xi$ is aligned (with the
Weyl tensor) if it is aligned with a Weyl principal bivector.
\end{definition}
After this definition, a corollary immediately follows from
proposition \ref{prop-al-1}:
\begin{corollary}
Let $\xi$ be a (real) Killing field in a type I spacetime, and let
$\{ \Gamma_i^j \}$ the (complex) connection 1-forms associated
with the principal bivectors of the Weyl tensor. The necessary and
sufficient condition for the Papapetrou field $\nabla \xi$ to be
aligned with the principal bivector ${\cal U}_i$ ($\nabla \xi =
\Omega {\cal U}_i + \tilde{\Omega} \tilde{\cal U}_i$) is $\xi$ to
be orthogonal to the two complex connection 1-forms $\Gamma_i^j$.
\end{corollary}
This result is independent of the Ricci tensor. Now we will
analyze the case of vacuum solutions in detail.

In a type I spacetime, the connection 1-forms associated with the
principal bivectors of the Weyl tensor must also be
$\xi$-invariant. This condition means that $\mbox{d}
(\xi,\Gamma_i^j) =-i(\xi) \mbox{d} \Gamma_i^j$. Moreover, if the
Ricci tensor is zero, we can use the second structure equations
(\ref{es2a}) with $Q=0$ to substitute the differential of the
connection 1-forms and we obtain that every Killing field $\xi$
must satisfy
\begin{equation} \label{ka5}
\mbox{d} \Omega_i = \Omega_j \ \Gamma_i^j + \alpha_i \ {\cal U}_i
(\xi)
\end{equation}
Thus, if the Killing 2--form $\nabla \xi$ is aligned with a
principal bivector, let us say ${\cal U}_3$, $\xi$ must be
orthogonal to $\Gamma_3^1$ and $\Gamma_3^2$ and so $\Omega_1
=0=\Omega_2$. In this particular case, equations (\ref{ka5})
become
\begin{equation} \label{ka6}
\hspace{-1cm} \mbox{d} \Omega_3=\alpha_3 \  {\cal U}_3(\xi) \, ,
\qquad \Omega_3 \ \Gamma_2^3  = - \alpha_2 \ {\cal U}_2 (\xi) \, ,
\qquad \Omega_3 \ \Gamma_1^3  = -\alpha_1 \ {\cal U}_1 (\xi)
\end{equation}
Taking into account that ${\cal U}^2_i = \frac{1}{2} g $, from the
second of the equations above we obtain $\xi = 2
\frac{\Omega_3}{\alpha_2} \ {\cal U}_2 (\Gamma_3^2)$. So, it
follows that if $\xi_1$ and $\xi_2$ are two Killing fields which
are orthogonal to the same pair of connection 1-forms, then $\xi_1
\wedge \xi_2 =0$. This result can be stated as:
\begin{proposition}
Let $\{ {\cal U}_i \}$ be the principal bivectors of a type I
vacuum solution. Then, for every ${\cal U}_i$, there is at most,
one (real) Killing field $\xi_i$ such that its associated
Papapetrou field $\nabla \xi_i$ is aligned with ${\cal U}_i$.
\end{proposition}
Equations (\ref{ka6}) can be written in the equivalent form
\begin{equation}\label{ka7}
\hspace{-1.5cm} \xi = \frac{2}{\alpha_3}  {\cal U}_3 (\mbox{d} \Omega_3)\, , \qquad
\Gamma_3^2=  \mbox{\rm i}  \sqrt{2}  \frac{\alpha_2}{\alpha_3}  {\cal U}_1 (\mbox{d}
 \ln{\Omega_3})\, , \qquad
\Gamma_1^3 = \mbox{\rm i} \sqrt{2} \frac{\alpha_1}{\alpha_3} {\cal
U}_2 (\mbox{d} \ln \Omega_3 )
\end{equation}
Taking into account these expressions and (\ref{05}), from  Bianchi identities
(\ref{bianchi}) we obtain
\begin{equation} \label{ka8}
\mbox{d} (\alpha_3)^2  = 4 ( \alpha_2^2 + \alpha_2 \alpha_3 +
\alpha_3^2 ) \ \mbox{d} \ln{\Omega_3}
\end{equation}
So $\mbox{d} \alpha_3 \wedge \mbox{d} \ln{\Omega_3} =0$, and if we
differentiate (\ref{ka8}) we have $\, (2 \alpha_2 + \alpha_3 )
\mbox{d} \alpha_2 \wedge \mbox{d} \ln{\Omega_3} = 0$. Then, as $(2
\alpha_2 + \alpha_3 ) \neq 0$ we conclude that
\begin{equation}\label{ka10}
\mbox{d} \alpha_2 \wedge \mbox{d} \ln{\Omega_3}  =0
\end{equation}
Moreover, if $\alpha_2^2 + \alpha_2 \alpha_3 + \alpha_3^2 =0$,
from (\ref{ka8}) we have $\mbox{d} \alpha_3 =0$ and so $\mbox{d}
\alpha_2 = 0$. Thus,  the eigenvalues are constant and the Bianchi
identities lead to $\lambda_i \wedge \lambda_j  =0$. So, the
spacetime belongs to class I$_1$. On the other hand, if
$\alpha_2^2 + \alpha_2 \alpha_3 + \alpha_3^2  \neq 0$, from
(\ref{ka8}) and (\ref{ka10}) we have $\mbox{d} \alpha_2 \wedge
\mbox{d} \alpha_3 =0$. But (\ref{ka7}) and (\ref{bianchi}) imply,
$$\mbox{d} \alpha_2 \wedge \mbox{d} \alpha_3 =
\frac{2}{\alpha_1} (\alpha_1 - \alpha_2 )( \alpha_2^2 + \alpha_2 \alpha_3 + \alpha_3^2 ) \
 {\cal U}_2(\Gamma_1^3 ) \wedge  {\cal U}_3(\Gamma_1^2) $$
and so, it follows that $\, {\cal U}_2(\Gamma_1^3 ) \wedge  {\cal
U}_3(\Gamma_1^2) =0$. From expressions (\ref{05}), this last
condition states that
 $\lambda_i \wedge \lambda_j =0$, and so we have established:
\begin{theorem} \label{theopap}
A vacuum type I spacetime which admits a Killing field with an
aligned Papapetrou field belongs to class I$_1$.
\end{theorem}
In following sections we study the symmetries of the class $I_1$
spacetimes and we will show that they admit more than one Killing
field. Consequently, a unique symmetry with an aligned Papapetrou
field implies that other symmetries exist. More precisely, from
theorem \ref{theopap} above and theorem \ref{theosym} and
proposition \ref{propet} that we will show in following sections,
we can state:

\begin{corollary}
A vacuum type I spacetime which admits a Killing field with an
aligned Papapetrou field admits, at least, a 3-dimensional group
of isometries.
\end{corollary}

\section{Type I$_1$ vacuum solutions: basic properties and classification}
In this section we analyze some basic properties of type I$_1$
vacuum metrics which lead us to a natural subclassification.
Afterwards in following sections, we study the symmetries of these
spacetimes. We shall start our analysis of type I$_1$ vacuum
metrics for the case of solutions having a non constant eigenvalue
$\alpha_1$. The case of all the eigenvalues being constant will be
dealt with in the last section.

Type I$_1$ vacuum solutions satisfy $\lambda_i \wedge \lambda_j
=0$. Then, if $\mbox{d} \alpha_1 \neq 0$, Bianchi identities
(\ref{bianchi}) show $\mbox{d} \alpha_1 \wedge \lambda_i =0$. Now
expressions (\ref{cap3123}) say that three functions $\gamma_i$
exist such that
 \begin{equation}\label{cero}
\Gamma_i^j = \gamma_k \ \epsilon_{ijk} \ {\cal U}_k ( \mbox{d}
\alpha_1)
\end{equation}
In this case, the second structure equations (\ref{es2a}) can be written as
\begin{equation} \label{es2}
\begin{array}{r}
\hspace*{-1.5 cm} \mbox{d} {\cal U}_i (\mbox{d} \alpha_1) = -
\mbox{d} \ln \gamma_i \wedge {\cal U}_i (\mbox{d} \alpha_1) -
\frac{\gamma_j \gamma_k}{\gamma_i} \ {\cal U}_j (\mbox{d} \alpha_1
) \wedge {\cal U}_k (\mbox{d} \alpha_1) + \ci \sqrt{2} \
\frac{\alpha_i}{\gamma_i} \ {\cal U}_i
\end{array}
\end{equation}
for every cyclic permutation $i,j,k$. In terms of these functions
$\gamma_i$,  the Bianchi identities (\ref{bianchi}) write:
\begin{equation} \label{34}
\begin{array}{l}
\mbox{d} \alpha_1 = \frac{1}{\sqrt{2}} \Big( \, \gamma_3 (\alpha_1
- \alpha_2 )
+ \gamma_2 (\alpha_1 - \alpha_3) \Big) \mbox{d} \alpha_1 \\[3mm]
\mbox{d} \alpha_2 = \frac{1}{\sqrt{2}} \Big( \, \gamma_3 (\alpha_2
- \alpha_1) + \gamma_1 (\alpha_2 - \alpha_3) \Big) \mbox{d}
\alpha_1
\end{array}
\end{equation}
The second equation above says that  $\alpha_2$ depends on
$\alpha_1$. Now we will   prove that  the functions  $\gamma_k$
depend on $\alpha_1$ too. From equations (\ref{cero}) we have  $\,
\Gamma_1^2 = \gamma_3 \ {\cal U}_3(\mbox{d} \alpha_1)$, and we can
calculate $\nabla \Gamma_1^2$ and make use of (\ref{es1a}) to
substitute $\nabla {\cal U}_3$. Thus, we  obtain
$$\nabla \Gamma_1^2 = \mbox{d} \ln \gamma_3 \otimes \Gamma_1^2 - \gamma_3
(\nabla \mbox{d} \alpha_1 ) \times {\cal U}_3 -
\frac{\gamma_3}{\gamma_1} \ \Gamma_1^3 \otimes \Gamma_2^3 +
\frac{\gamma_3}{\gamma_2} \ \Gamma_2^3 \otimes \Gamma_1^3 $$
Contracting this equation with ${\cal U}_1$ and ${\cal U}_2$ and
taking into account (\ref{cero}) and that $\nabla \mbox{d}
\alpha_1$ is symmetric, we get
$${\cal U}_j (\Gamma_1^2 , \mbox{d} \ln \gamma_3 ) = -({\cal U}_j ,
\mbox{d} \Gamma_1^2) = 0 \, \qquad  j=1,2$$ where, in the last
equality, we have used the second structure equations (\ref{es2})
and expressions (\ref{cero}). So we find that $\mbox{d} \gamma_3$
is orthogonal to ${\cal U}_1 (\Gamma_1^2)$ and ${\cal U}_2
(\Gamma_1^2)$, or from (\ref{cero}), $\mbox{d} \gamma_3$ is
orthogonal to ${\cal U}_1 (\mbox{d}{\alpha_1})$ and ${\cal U}_2 (
\mbox{d} \alpha_1)$. If we repeat this calculation  replacing
$\Gamma_1^2$ by $\Gamma_1^3$, we also see that $\mbox{d} \gamma_2$
is orthogonal to ${\cal U}_3(\mbox{d} \alpha_1)$ and ${\cal U}_1
(\mbox{d} \alpha_1)$. According to (\ref{34}) $\gamma_3$  depends
on $\alpha_1$ and $\gamma_2$, so we have that $\mbox{d} \gamma_3$
is orthogonal to every ${\cal U}_j (\mbox{d} \alpha_1)$. Equations
(\ref{34}) say that the same applies for $\mbox{d} \gamma_2$ and
$\mbox{d} \gamma_1$. Thus we have established that, for all $i,j$,
\begin{equation} \label{dgamma}
(\mbox{d} \gamma_i , {\cal U}_j (\mbox{d} \alpha_1) ) =0
\end{equation}
If $\mbox{d} \alpha_1 $ is not a null vector, $\{ \mbox{d}
\alpha_1 , {\cal U}_i (\mbox{d}{\alpha_1} ) \}$ is an orthogonal
frame and (\ref{dgamma}) proves that $\mbox{d} \gamma_j \wedge
\mbox{d} \alpha_1 =0$. So, we can state:

\begin{lemma} \label{nueva1}
Let $g$ be a class I$_1$ vacuum  solution with the Weyl tensor
having an eigenvalue $\alpha_1$ such that $\, ({\mbox{\rm d}
\alpha_1})^2 \not=0$. Then the functions $\gamma_j$ given in
(\ref{cero}) satisfy $\, \mbox{\rm d} \gamma_j \wedge \mbox{\rm d}
\alpha_1 =0$.
\end{lemma}

On the other hand, the second structure equations (\ref{es2}) give
us the differential of the 1-forms ${\cal U}_i (\mbox{d}
\alpha_1)$. If we consider an arbitrary (complex) vector field
$\chi$ and the 1-forms $\{ {\cal U}_i (\chi) \}$, expressions
(\ref{orient}) and (\ref{es1a}) can be used to compute the Lie
derivative ${\cal L}_{\chi} g$ in terms of the exterior
differentials $\{ \mbox{d} {\cal U}_i (\chi) \}$ and the
connection coefficients associated with $\{ {\cal U}_i \}$. More
precisely, we have:
\begin{equation} \label{lie}
{\cal L}_{\chi} g \equiv  ( \nabla \chi  +  ^{t}\nabla \chi) = 2 \
({\cal U}_1 \times \Lambda_{1} +  \Lambda_{2} \times {\cal U}_2 +
i \sqrt{2}\, {\cal U}_1 \times \Lambda_3 \times {\cal U}_2 )
\end{equation}
where $\, (\Lambda_i)_{\alpha \beta} = -\mbox{d}( {\cal U}_i (\chi
))_{\alpha \beta} - \chi^{\epsilon} \left( \nabla_{\alpha} {({\cal
U}_i)}_{\epsilon\beta} - \nabla_{\beta} {({\cal
U}_i)}_{\epsilon\alpha} \right)$. As $\nabla \mbox{d} \alpha_1$ is
a symmetric tensor and  $\, \mbox{d} (\mbox{d} \alpha_1 )^2 = 2
(\nabla \mbox{d} \alpha_1, \mbox{d} \alpha_1)$, we can use
expression (\ref{lie}) with $\chi = \mbox{d} \alpha_1$ and the
second structure equations (\ref{es2}) to compute $\mbox{d}
(\mbox{d} \alpha_1 )^2 $. Thus we obtain:
\begin{equation} \label{quadrat}
\begin{array}{lll}
\mbox{d} (\mbox{d} \alpha_1 )^2 =  & - &
\frac{1}{\gamma_3}(\mbox{d} \alpha_1)^2 \Big( \mbox{d} \gamma_3 +
\frac{1}{2} (\gamma_1 \gamma_2 - \gamma_2 \gamma_3 -
\gamma_1 \gamma_3) \, \mbox{d} \alpha_1 \Big) \ - \\[3mm]
 & - & ( \mbox{d} \ln \gamma_1 , \mbox{d} \alpha_1 )
 \ \mbox{d} \alpha_1  + \ci \sqrt{2} ( \frac{\alpha_3}{\gamma_3} -
\frac{\alpha_1}{\gamma_1} - \frac{\alpha_2}{\gamma_2})  \,
\mbox{d} \alpha_1
\end{array}
\end{equation}
If $\mbox{d} \alpha_1 $ is not a null vector lemma \ref{nueva1}
applies and $\mbox{d} \gamma_3 \wedge \mbox{d} \alpha_1 =0$. Then,
from (\ref{quadrat}) we have:

\begin{lemma} \label{nueva2}
Let $g$ be a class I$_1$ vacuum  solution with the Weyl tensor
having an eigenvalue $\alpha_1$ such that $\, ({\mbox{\rm d}
\alpha_1})^2 \not=0$. Then the function $(\mbox{\rm d} \alpha_1
)^2$ satisfies $\, \mbox{\rm d} (\mbox{\rm d} \alpha_1 )^2 \wedge
\mbox{\rm d} \alpha_1 =0$.
\end{lemma}

Finally we will prove now that $\mbox{d} \alpha_1$ can not be a
null vector. Let us suppose $({\mbox{d} \alpha_1})^2 =0$. Then,
from lemma \ref{is}, $\mbox{d} \alpha_1$ would be a combination of
the vectors ${\cal U}_i (\mbox{d} \alpha_1)$, and from
(\ref{dgamma}) we have:
\begin{equation} \label{36}
(\mbox{d} \alpha_1 , \mbox{d} \gamma_i ) =0
\end{equation}
Using (\ref{05}) to express  $\lambda_i$ in terms of $\gamma_i$
and $\mbox{d} \alpha_1$, the equations (\ref{traza}) for the
vacuum case  can be written as:
$$- \frac{\ci}{\sqrt{2}}  \ ( \mbox{d} (\gamma_j + \gamma_k), \mbox{d} \alpha_1 ) -
\frac{\ci}{\sqrt{2}} \ (\gamma_j + \gamma_k ) \ \Delta \alpha_1 +
\gamma_j \gamma_k (\mbox{d} \alpha_1)^2 = 2 \alpha_i $$ for every
cyclic permutation of $i,j,k$. Now, if $(\mbox{d} \alpha_1)^2 =0$
and taking into account (\ref{36}), we have $\, (\gamma_j +
\gamma_k) \Delta \alpha_1 = \ci 2 \sqrt{2} \ \alpha_i $. Solving
these equations we obtain:
\begin{equation} \label{gamma}
\gamma_2 = \frac{\alpha_2}{\alpha_3} \ \gamma_3 ; \ \ \ \gamma_1 =
\frac{\alpha_1}{\alpha_3} \ \gamma_3
\end{equation}
On the other hand, taking into account (\ref{36}) and $\,
(\mbox{d} \alpha_1)^2 =0$, equation (\ref{quadrat}) says that $\,
\frac{\alpha_3}{\gamma_3}-\frac{\alpha_1}{\gamma_1} -
\frac{\alpha_2}{\gamma_2}  =0$. But  from (\ref{gamma}) we arrive
to $\, \alpha_3 =0$, which is not compatible with the vacuum
condition as a consequence of the Szekeres-Brans theorem.
Therefore, $\mbox{d} \alpha_1$ can not be a null vector. We
summarize these results and lemmas \ref{nueva1} and \ref{nueva2}
in the following
\begin{proposition}  \label{nonul}
Let  $g$ be a class I$_1$ vacuum  solution with a non constant
Weyl eigenvalue $\alpha_1$. Then, it holds:

i) $(\mbox{\rm d} \alpha_1 )^2 \not=0$

ii) $\mbox{\rm d} (\mbox{\rm d} \alpha_1)^2 \wedge \mbox{\rm d}
\alpha_1 =0 $.

iii) $\mbox{\rm d} \gamma_k \wedge \mbox{\rm d} \alpha_1 = 0$
where $\gamma_k$ are the functions given in (\ref{cero}).
 \end{proposition}

The properties of the vacuum solutions of class $I_1$ summarized
in proposition \ref{nonul} give us the basic elements to analyze
this family of metrics when a non constant eigenvalue $\alpha_1$
exists. Indeed, being $\mbox{d} \alpha_1$ a non null vector, we
can use (\ref{ui}) with $x = \mbox{d} \alpha_1$ to eliminate $\,
{\cal U}_i$ from the second structure equations (\ref{es2}) and we
obtain  that those equations become:
\begin{equation}\label{es2def}
\mbox{d} {\cal U}_i (\mbox{d} \alpha_1) = \bar{\mu_i} (\alpha_1) \
\mbox{d} \alpha_1 \wedge {\cal U}_i (\mbox{d} \alpha_1) +
\bar{\nu_i} (\alpha_1) \ {\cal U}_j (\mbox{d} \alpha_1) \wedge
{\cal U}_k (\mbox{d} \alpha_1)
\end{equation}
where the functions $\bar{\mu_i}$ and $\bar{\nu_i}$ are given by:
$$
\bar{\mu_i} = \bar{\mu_i} (\alpha_1) \equiv - \frac{\mbox{d} \ln
\gamma_i}{\mbox{d} \alpha_1} + \frac{\sqrt{2} \alpha_i}{\gamma_i \
(\mbox{d} \alpha_1)^2 } ; \quad \quad \bar{\nu_i} =  \bar{\nu_i}
(\alpha_1) \equiv- \ci \Big( \frac{2 \alpha_i}{\gamma_i \
(\mbox{d} \alpha_1)^2} + \frac{ \gamma_j \gamma_k }{\gamma_i}
\Big)
$$
If ${\cal U}_i (\mbox{d} \alpha_1)$ satisfy the second structure
equations (\ref{es2def}), we can use (\ref{metrica1}) to find the
metric tensor as
\begin{equation}\label{metrica}
g = \frac{1}{(\mbox{d} \alpha_1)^2} \Big( \mbox{d} \alpha_1
\otimes \mbox{d} \alpha_1 - 2 \ \sum {\cal U}_i (\mbox{d}
\alpha_1) \otimes {\cal U}_i (\mbox{d} \alpha_1) \Big)
\end{equation}
This seems a hard task, not because of the procedure, but because
real coordinates must be adapted to the complex 1-forms ${\cal
U}_i(\mbox{d} \alpha_1) $. In this work we do not go on to the
explicit integration of the vacuum Einstein equation, but some
results of this kind will be presented elsewhere \cite{fsI1b}. At
this point, it is clear that the integration of the system
(\ref{es2def}) and, consequently, the gravitational field which is
solution of it, depends strongly on the number of the ${\cal
U}_i(\mbox{d} \alpha_1) $ that are integrable 1--forms. We will
see in the next section that this condition determines the group
of isometries of the spacetime. So it seems suitable to give an
invariant classification of type $I_1$ spacetimes that takes into
account these restrictions:
\begin{definition} \label{subclass}
We will say that a type I$_1$  vacuum metric with $\mbox{\rm d}
\alpha_1 \neq 0$ is of class I$_{1A}$ ($A=0,1,2,3$) if there are
exactly $A$ integrable 1--forms in the set $\{ {\cal U}_i
(\mbox{\rm d} \alpha_1) \}$.
\end{definition}

\section{Class I$_1$ vacuum solutions with non constant eigenvalues: symmetries}

Let us consider now a type $I$ metric that admits a Weyl
eigenvalue $\alpha_1$ with non null gradient, $(\mbox{d}
\alpha_1)^2 \neq 0$. The orthogonal frame $\{ \mbox{d} \alpha_1 ,
{\cal U}_{i} (\mbox{d} \alpha_1 ) \} $ is built up with invariants
and so these 1--forms and their square $(\mbox{d} \alpha_1)^2$
must be invariant with respect to every Killing field $\xi$. On
the other hand, if $\xi$ is a vector field such that $\mbox{d}
\alpha_1$, ${\cal U}_{i} (\mbox{d} \alpha_1) $ and $(\mbox{d}
\alpha_1)^2$ are $\xi$--invariant, then it must be a Killing field
because of (\ref{metrica}). Thus, we arrive to the following
result:
\begin{lemma} \label{k1}
Let $g$ be a type I metric such that an eigenvalue of the Weyl
tensor exists satisfying $(\mbox{\rm d} \alpha_1)^2 \neq 0 $, and
let ${\cal U}_i$ be the principal bivectors. Then,  $\xi$ is a
Killing field if, and only if, it satisfies
$${\cal L}_{\xi} \mbox{\rm d} \alpha_1 = {\cal L}_{\xi} {\cal U}_i(\mbox{\rm d} \alpha_1)=
{\cal L}_{\xi} (\mbox{\rm d} \alpha_1)^2 =0$$
\end{lemma}

In the case of a vacuum class I$_1$ spacetime with a non constant
eigenvalue $\alpha_1$, the function $(\mbox{d} \alpha_1)^2 \neq 0$
depends on $\alpha_1$ as proposition \ref{nonul} states. So, the
vector fields orthogonal to $\mbox{d} \alpha_1$ are those that
leave invariant the scalar $(\mbox{d} \alpha_1)^2$. But as the
space orthogonal to $\mbox{d} \alpha_1$ is generated by the
vectors ${\cal U}_i (\mbox{d} \alpha_1)$, every Killing field
$\xi$ must be a combination of them. Then, lemma \ref{k1} can be
stated for the $I_1$ case as:
\begin{proposition} \label{k2}
Let $g$ be a class I$_1$ vacuum metric such that $\mbox{\rm d}
\alpha_1 \neq 0$. A vector $\xi$ is a Killing field if, and only
if, it satisfies:

(i) $\xi \wedge {\cal U}_1 (\mbox{\rm d} \alpha_1) \wedge {\cal
U}_2 (\mbox{\rm d} \alpha_1)  \wedge {\cal U}_3 (\mbox{\rm d}
\alpha_1) =0$.

(ii) ${\cal L}_{\xi} {\cal U}_i (\mbox{\rm d} \alpha_1) = 0$.
\end{proposition}
If we consider three functions depending on $\alpha_1$,
$m_i(\alpha_1)$, then the 1-forms $M_i = m_i (\alpha_1) \ {\cal
U}_i(\mbox{d} \alpha_1) $ also satisfy the conditions (i) and (ii)
of the proposition \ref{k2}. Moreover $m_i(\alpha_1)$ exist  such
that the equations (\ref{es2def}) write for the 1-forms $M_i = m_i
(\alpha_1) \ {\cal U}_i(\mbox{d} \alpha_1) $ as the exterior
system
\begin{equation} \label{5}
\mbox{d}{M_1} = \delta_1 M_2 \wedge M_3 ; \  \ \ \
  \mbox{d}{M_2} = \delta_2 M_3 \wedge M_1 ; \  \ \ \
\mbox{d}{M_3} = \delta_3 M_2 \wedge M_1
\end{equation}
where $\delta_i$ takes the value $0$ if ${\cal U}_i (\mbox{d}
\alpha_1)$ is integrable and value $1$ if this does not hold.
Thus, the 1-forms $M_i = m_i (\alpha_1) \ {\cal U}_i(\mbox{d}
\alpha_1) $ can be considered as the dual 1-forms of the
reciprocal group of a transitive group $G_3$ of isometries.
Moreover, the (complex) Bianchi type will depend on the integrable
character of every ${\cal U}_1 (\mbox{d} \alpha_1)$. Thus, taking
into account definition \ref{subclass} we have:

\begin{theorem} \label{theosym}
The class $I_1$ vacuum solutions with a non constant Weyl
eigenvalue admit a $G_3$ group of isometries. The Bianchi type
depends on the subclasses $I_{1A}$.
\end{theorem}

These results allow us to analyze in detail the Bianchi type of
every class $I_{1A}$ and to study when a Killing 2--form can be
aligned with the Weyl tensor.

\subsection{Class I$_{10}$ vacuum solutions}

If none of the directions $M_i$ is vorticity free, we can choose
every $\delta_i$ of (\ref{5}) to take value $1$. So, complex
coordinates exist such that the 1--forms $M_i$ take the canonical
expression of the reciprocal group of the (complex) Bianchi type
VIII that corresponds to the real types VIII and IX \cite{kra}.
But the system can also be integrated in complex coordinates to
get
\begin{equation} \label{b8}
\begin{array}{l}
M_1 = - \frac{\ci}{2} [ \ e^x \mbox{d} z + e^{-x} (2 \mbox{d} y - y^2 \mbox{d} z) ] \\
M_2 = \frac{1}{2} [ \ e^x \mbox{d} z - e^{-x} (2 \mbox{d} y - y^2 \mbox{d} z) ] \\
M_3 = \ci (\mbox{d} x - y \mbox{d} z )
\end{array}
\end{equation}
In this coordinate system the (complex) Killing fields can be expressed as
\begin{equation} \label{b8k}
\xi = (k_2 + 2 k_1 z) \partial_x + (y(2k_1 + k_2)-2k_1) \partial_y
+(k_1 z^2 + k_2 z + k_3) \partial_z
\end{equation}

To see if an aligned Killing 2--form exists we must impose a
Killing field to be orthogonal to two connection $1$-forms. As
every connection 1-form takes the direction of one of the $M_i$,
we must see if there is a Killing field which is orthogonal to two
of the 1-forms $M_i$. But from the general expression of a Killing
field (\ref{b8k}) if $\xi$ is orthogonal to $M_3$ then $\xi=0$.
The only possibility is  $\xi$ to be orthogonal to $M_1$ and
$M_2$. But as $e^x$ and $e^{-x}$ are independent, we also obtain
$\, \xi=0$. So, we can conclude:
\begin{proposition}
Every vacuum solution of class I$_{10}$  admits a G$_3$ of Bianchi type VIII or IX.
 In such spacetime there is no Killing 2-form aligned with the Weyl tensor.
\end{proposition}

\subsection{Class I$_{11}$ vacuum solutions}

Let us suppose that one of the 1-forms $M_i$, say $M_2$ is
integrable. Then, the exterior system (\ref{5}) becomes the system
satisfied by the dual 1-forms of the reciprocal group of the
(complex) Bianchi type VI which corresponds to the real types VI
or VII. Both cases can be integrated at once in complex
coordinates $\{ x, y , z \}$ to obtain
\begin{equation}
M_1 + M_3 = e^{-z} \mbox{d} x ; \qquad M_1 - M_3 = e^{z } \mbox{d}
y , \qquad M_2 = \mbox{d} z
\end{equation}
In this coordinate system, the field
\begin{equation}
\xi = (k_1 + k_3 x)  \partial_x + (k_2 - k_3 y ) \partial_y + k_3
\partial_z
\end{equation}
is a (complex) Killing field for arbitrary values of the constants $k_i$.

To see if there is a Killing field with an aligned Killing
2-form, we must impose a Killing field to be orthogonal to two of the
1-forms $M_i$. A straightforward calculation
shows that in this case
$$\begin{array}{l}
(\xi , M_2 ) = k_3 \\[3mm]
(\xi , M_1 )= (k_1 + k_3 x )    e^{-z} + (k_2 - k_3 y) e^{z} \\[3mm]
(\xi , M_3) = (k_1 + k_3 x) e^{-z} - (k_2 - k_3 y ) e^z \\[3mm]
\end{array}$$
From here it is easy to show that there is no Killing field that
is orthogonal to two of the connection 1-forms. So, we have shown:
\begin{proposition} Every vacuum solution of class I$_{11}$ admits
a G$_3$ of Bianchi type VI or VII. In such spacetime there is
no Killing 2-form aligned with the Weyl tensor.
\end{proposition}

\subsection{Class I$_{12}$ vacuum solutions}

Let us suppose that $M_1$ and $M_2$ are integrable. The exterior system (\ref{5})
can be easily integrated in complex coordinates  to get
\begin{equation}
M_1  = \mbox{d} x , \qquad M_2 = \mbox{d} y , \qquad M_3  = - x
\mbox{d} y + \mbox{d} z
\end{equation}
which correspond to the reciprocal group of a Bianchi type II.
Moreover, for arbitrary values of the constants $k_i$ the field
 $$\xi = k_1 \partial_x + k_2 \partial_y + (k_1 y + k_3 ) \partial_z $$
is a (complex) Killing field.

A straightforward calculation shows that the only Killing field
that is orthogonal to two connection 1-forms is $\xi=\partial_z$.
But as $z$ is a complex coordinate we can not conclude that it
defines a unique real Killing field and so we can not ensure that
an aligned (real) Killing 2-form exits. Thus, at this point we can
state:
\begin{proposition}
Every vacuum solution of class I$_{12}$ admits a G$_3$ of Bianchi
type II. If ${\cal U}_j(\mbox{\rm d} \alpha_1)$ is the unique non
integrable 1--form, then ${\cal U}_j$ is the only principal
bivector that could be aligned with a Killing 2--form.
\end{proposition}

\subsection{Class I$_{13}$ vacuum solutions}
If all of the 1-forms $M_i$ are integrable, the system (\ref{5})
says that three complex functions $\{ x_i \}$ exist, such that
\begin{equation}
M_i = \mbox{d} x_i
\end{equation}
This corresponds to a commutative group G$_3$ of isometries.

Moreover, for every pair of connection 1-forms a complex Killing
field exits such that its Killing 2-form is aligned with it. But
as happened before we can not conclude here that a real aligned
Killing 2-form exits. At this point we can state:
\begin{proposition}
Every vacuum solution of class I$_{13}$ admits a G$_3$ of Bianchi
type I. Every principal bivector ${\cal U}_i$ could be aligned
with a Killing 2--form.
\end{proposition}

\section{Type I vacuum solutions with constant eigenvalues}

If a type $I$ metric satisfies $\mbox{d} \alpha_i =0$, the Bianchi
identities (\ref{bianchi}) imply
\begin{equation}\label{petrov1}
\lambda_2 = \frac{(2 \alpha_1 + \alpha_2)^2}{(\alpha_1 + 2 \alpha_2
)^2} \ \lambda_1 \, , \qquad \quad  \lambda_3 = \frac{(\alpha_1 -
\alpha_2)^2}{(\alpha_1 + 2 \alpha_2 )^2} \ \lambda_1
\end{equation}
and so we have that the metric always belongs to class I$_1$. This
fact can be stated as
\begin{proposition}
Every type I spacetime with vanishing Cotton tensor and constant
Weyl eigenvalues is of class I$_1$.
\end{proposition}
Using (\ref{cap3123}), conditions (\ref{petrov1})
can be stated in terms of the connection 1-forms as
\begin{equation} \label{pet5}
\Gamma_1^3 = -\ci \sqrt{2} \ \frac{\alpha_1 - \alpha_2}{\alpha_1
-\alpha_3} \, {\cal U}_1 (\Gamma_1^2) ; \quad \quad \Gamma_2^3 =
-\ci \sqrt{2} \ \frac{\alpha_2 - \alpha_1}{\alpha_2 - \alpha_3} \,
{\cal U}_2(\Gamma_1^2)
\end{equation}
Putting (\ref{petrov1}) in (\ref{traza}) with $Q=0$, we obtain
\begin{equation}\label{petrov2}
{\alpha_1}^2  + \alpha_1 \alpha_2 + {\alpha_2}^2 =0 , \qquad
(\lambda_1 , \lambda_1 ) = -\frac{ \alpha_1}{4}, \qquad
\nabla \cdot \lambda_1 =0
\end{equation}
The first of these equations states that $\tr {\cal W}^2 =0$ and so,
the four Debever vectors define a symmetric frame as was established in \cite{fsEM}.
Moreover, as $(\lambda_1 , \lambda_1 ) = -\frac{ \alpha_1}{4}$ and using (\ref{05}) we
deduce
$$(\Gamma_1^2)^2 = - \frac{{\alpha_2}^2}{\alpha_1} = - \alpha_3 $$
where the first expression of (\ref{petrov2}) has been used to
eliminate ${\alpha_2}^2$. So, $(\Gamma_1^2)^2 \neq 0$, and we can
consider the orthogonal frame $\{ \Gamma_1^2 , {\cal U}_i
(\Gamma_1^2)  \}$, where every 1--form has, up to a factor 2, the
same constant modulus $\sqrt{-\alpha_3}$. Taking into account
(\ref{pet5}), the second structure equations (\ref{es2a}) with
$Q=0$ just write
\begin{equation}\label{pet6}
\begin{array}{l}
\mbox{d} \Gamma_1^2 = \ci \sqrt{2} \ \Gamma_1^2 \wedge {\cal U}_3 (\Gamma_1^2) \\[3mm]
\mbox{d} {\cal U}_1 (\Gamma_1^2) = \ci \sqrt{2} \
\frac{\alpha_3}{\alpha_2} \
{\cal U}_1 (\Gamma_1^2) \wedge {\cal U}_3 (\Gamma_1^2) \\[3mm]
\mbox{d} {\cal U}_2 (\Gamma_1^2) = \ci \sqrt{2} \
\frac{\alpha_1}{\alpha_2} \ {\cal U}_2 (\Gamma_1^2) \wedge {\cal
U}_3 (\Gamma_1^2)
\end{array}
\end{equation}
As the Weyl eigenvalues are constant, the integrability conditions
of (\ref{pet6}) state $\mbox{d} {\cal U}_3(\Gamma_1^2)  =0$. Then,
the exterior system can easily be integrated in complex
coordinates $\{ x, y , z, w \} $ to obtain
\begin{equation}\label{solpet}
\begin{array}{l}
\displaystyle \Gamma_1^2  = e^{-\ci  \sqrt{2} w} \mbox{d} x \\[3mm]
\displaystyle   {\cal U}_1 (\Gamma_1^2)  = e^{- \ci \sqrt{2}
\frac{\alpha_3}{\alpha_2} w}
\mbox{d} y \\[3mm]
\displaystyle  {\cal U}_2 (\Gamma_1^2) = e^{- \ci \sqrt{2}
\frac{\alpha_1}{\alpha_2} w} \mbox{d} z  \\[3mm]
 {\cal U}_3(\Gamma_1^2)   = \mbox{d} w
\end{array}
\end{equation}
The 1--forms (\ref{solpet}) define an orthogonal frame built up
with invariants. So, $\xi$ is a Killing field if, and only if, it
leaves the frame unchanged, ${\cal L}_{\xi} \Gamma_1^2  =0$,
${\cal L}_{\xi} {\cal U}_i (\Gamma_1^2) =0$. If we write $\xi$ as
a linear combination of the coordinate fields $\{ \partial_{x},
\partial_y,
\partial_z , \partial_w \}$, we find that for arbitrary values of
the constants $k_i$, the fields
\begin{equation}
\hspace{-2cm} \xi = k_4 \partial_w +
(- \mbox{\rm i} \sqrt{2} k_4 x + k_1 ) \partial_x +
(- \mbox{\rm i} \sqrt{2} \frac{\alpha_3}{\alpha_2} k_4 y + k_2 )
\partial_y
+ (- \mbox{\rm i} \sqrt{2} \frac{\alpha_1}{\alpha_2} k_4 z + k_3 ) \partial_z
\end{equation}
are Killing fields, and so, a G$_4$ exists. Then, the spacetimes
is the Petrov homogeneous vacuum solution \cite{pet} \cite{kra}:

\begin{proposition} \label{propet}
The only Type I vacuum solution with Weyl constant eigenvalues is
the Petrov solution, and so it admits a four dimensional group of
isometries.
\end{proposition}
This proposition provides a {\it intrinsic} (depending solely on the
 metric tensor) characterization of the Petrov homogeneous vacuum solution.

Indeed, the Petrov solution can be found as the only one
satisfying \cite{pet}: (i) vacuum, (ii) existence of a simply
transitive group $G_4$ of isometries. The first condition is
intrinsic because it imposes a restriction on a metric
concomitant, the Ricci tensor. Nevertheless, the second one
imposes equations that mix up, in principle, elements other than
the metric tensor (Killing vectors of the isometry group).
Proposition \ref{propet} substitutes this last non intrinsic
condition for an intrinsic one: the Weyl tensor is Petrov type I
with constant eigenvalues. Moreover, the characterization can also
become {\it explicit} because the metric concomitants admit known
explicit expressions in terms of the metric tensor \cite{fms}.
More precisely, if we consider that Weyl constant eigenvalues is
equivalent to constant Weyl symmetric scalars, we have:
\begin{theorem} \label{theopet}
Let $Ric(g)$ and ${\cal W} \equiv {\cal W}(g)$ the Ricci and the
self-dual Weyl tensor of a spacetime metric $g$. The necessary and
sufficient conditions for $g$ to be the Petrov homogeneous vacuum
solution are
\begin{equation}
Ric(g) = 0 \, ,  \qquad  (\tr{\cal W}^2)^3 \neq 6(\tr {\cal
W}^3)^2 \, , \qquad \mbox{d} \tr {\cal W}^2 = \mbox{d} \tr {\cal
W}^3 = 0
\end{equation}
\end{theorem}

If we want to study the alignment of the Killing 2--forms with the
Weyl tensor, we can compute the product of the Killing tensors
with the connection 1-forms taking into account (\ref{solpet}) and
(\ref{pet5}). Thus, we obtain
$$\begin{array}{l}
(\xi, \Gamma_1^2 ) = 0 \Longleftrightarrow - \ci \sqrt{2} k_4 x + k_1=0 \\[3mm]
(\xi, \Gamma_1^3 ) = 0  \Longleftrightarrow  - \ci \sqrt{2}
\frac{\alpha_3}{\alpha_2} k_4 y + k_2  =0\\[3mm]
(\xi, \Gamma_2^3 )= 0 \Longleftrightarrow - \ci \sqrt{2}
\frac{\alpha_1}{\alpha_2} k_4 z + k_3 =0
\end{array}$$
So, for every pair of connection 1-forms there is a Killing field
which is orthogonal to them. In order to see if an aligned Killing
2--form exists, we should have to prove if the  complex Killing
field defines only a real one. At this point we can state:

\begin{proposition}
The Petrov homogeneous vacuum solution could admit a Killing
2--form aligned with every Weyl principal bivector.
\end{proposition}

\section{Concluding remarks}

The results in this work show that the alignment of the Papapetrou
field associated with the Killing vector of a type I vacuum
solution with an isometry imposes strong complementary
restrictions on the metric tensor, namely, it admits a $G_3$ or a
$G_4$ group of isometries. On the contrary, we know \cite{fsDB}
that in the type D vacuum case the Kerr-NUT family has a
Papapetrou field aligned with the Weyl principal 2--planes, and
this family admits only a $G_2$, the minimum group of isometries
of a type D vacuum solution.

Our study is based in showing that the type I vacuum metrics with
an aligned Papapetrou field belong to the more degenerate class
$I_1$ of an invariant classification of type I spacetimes. All
these metrics admit at least a group $G_3$ of isometries and the
Bianchi type can also be characterized in terms of invariant
conditions imposed on the Weyl tensor. The full integration of the
vacuum equations is an ongoing study which will be presented
elsewhere \cite{fsI1b}. We are obtaining some known spatially
homogeneous vacuum solutions, like the Kasner or Taub metrics
\cite{kra}, as well as, their counterpart with time-like orbits.
Some solutions with time-like orbits which are not orthogonal to a
Weyl principal direction have also been found. The explicit
expression of these metrics in a coordinate system is necessary in
order to complete the results obtained here on the type I metrics
with aligned Papapetrou fields.

We have shown here that a type I vacuum metric with Weyl constant
eigenvalues admits a group $G_4$ of isometries and, consequently,
it is the Petrov homogeneous vacuum solution \cite{pet}. This
result allows us to give an intrinsic and explicit identification
of the Petrov solution in theorem \ref{theopet}. Elsewhere
\cite{fsS} we have pointed out the interest in obtaining a fully
intrinsic and explicit characterization of a metric or a family of
metrics. We have also explained the role that the covariant
determination of the Ricci and Weyl eigenvalues and eigenvectors
plays in this task \cite{fms}. In the natural sequel to the
present article \cite{fsI1b} we will integrate vacuum equations by
using a method which allows us to label every solution. In this
way we will obtain an intrinsic and explicit algorithm to identify
every type I vacuum metric admitting an aligned Papapetrou field.

\ack The authors would like thank A. Barnes and J.M.M. Senovilla
for bringing to light some references. This work has been
partially supported by the Spanish Ministerio de Ciencia y
Tecnolog\'{\i}a, project AYA2000-2045.

\end{document}